\documentclass[12pt, twoside]{article}

\usepackage{graphicx}
\usepackage{hyperref}

\begin{document}

\newtheorem{Theorem}{Theorem}

\begin{center}
{\Large\bf Coordinate Geometric Generalization of the Spherometer and Cylindrometer} \\

~\\

\noindent
Sameen Ahmed Khan \\
Engineering Department \\
Salalah College of Technology ({\bf SCOT}) \\
Post Box No. 608, Postal Code: 211 \\
Salalah, {\bf Sultanate of Oman}. \\
rohelakhan@yahoo.com, 
\url{http://SameenAhmedKhan.webs.com/} \\
\end{center}

\bigskip

\begin{abstract}
Spherometer is an instrument widely used for measuring the radius of curvature of a spherical surface. 
Cylindrometer is a modified spherometer, which can measure the radii of both spherical and 
cylindrical surfaces.  Both of these instruments are based on a geometric relation unique 
to circles and spheres, from Euclidean geometry.  A more general understanding is obtained 
using coordinate geometry.  The coordinate geometric approach also enables a generalization 
of the spherometer and cylindrometer to devices, which can handle aspherical surfaces.  
Here, we present the newly developed coordinate geometric approach and its applications.   
\end{abstract}



\noindent
{\bf OCIS:} 000.3860; 220.4610. \\

\noindent
{\bf Subj-class:} MP-Mathematical Physics \\
 

\newpage

{

\footnotesize


\baselineskip11pt

\tableofcontents

\vspace{-0.2cm}

\listoffigures


}

\newpage


\section{Introduction}
Spherometers are precision instruments designed to measure the radius of curvature of spherical 
surfaces as the name suggests.  They are particularly useful for situations, where only a 
portion of the spherical surface is available.
Such situations are very common in the optics workshops, while fabricating lenses 
and mirrors~\cite{Introductory-1}-\cite{Khan-lab}.  
They were designed by the opticians of the early-nineteenth century (or even earlier). 

\begin{quote}
{\em Although there is some thought that the spherometer was invented by the French 
optician Laroue, the first spherometer of which there is positive knowledge was devised and 
named around 1810 by Robert-Agla\'{e} Cauchoix, and made by the French mechanician Nicolas Fortin.  
Cauchoix's design, a three-legged base supporting a central micrometer screw, was quickly adopted 
as the basic standard and remains in use to this day.  
The {\em Conservatorie National des Arts et Metiers} in Paris has a spherometer made by Cauchoix and 
used by Biot that reads to 1/1000 millimeter}~\cite{Bud-Warner}. 
\end{quote}

\begin{figure}[ht]
\begin{center}
\includegraphics[width=8cm]{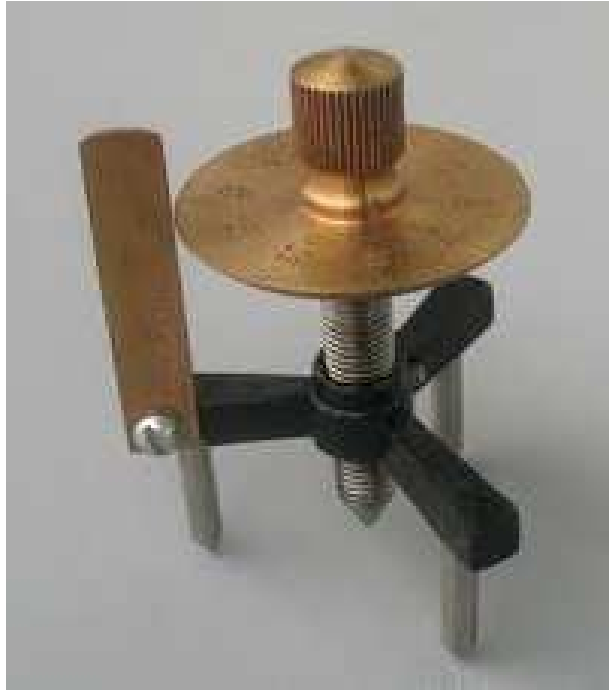} 
\caption{The Common Spherometer.}
\label{figure-common-spherometer}
\end{center}
\end{figure}

Figure-\ref{figure-common-spherometer} has the very widely used tripod design.
If the radii of both surfaces of the lens and the index of refraction of the lens material 
are known, the focal length can be found using the Lens-Maker's equation~\cite{Anderson-Burge}.
\begin{eqnarray}
P = \frac{1}{f} = (\mu - 1) \left[\frac{1}{R_1} - \frac{1}{R_2} + \frac{(\mu - 1) d}{\mu R_1 R_2} \right]\,, 
\label{lens-makers-equation}
\end{eqnarray}
where $P$ is the power of the lens, $f$ is the focal length of the lens, 
$\mu$ is the refractive index of the lens material, $R_1$ and $R_2$ are the radii of curvature 
of the lens surfaces and $d$ is the thickness of the lens.  In the thin lens approximation, 
we have  
\begin{eqnarray}
P = \frac{1}{f} \approx (\mu - 1) \left[\frac{1}{R_1} - \frac{1}{R_2} \right]\,.
\label{thin-lens-makers-equation}
\end{eqnarray}
The error arising from the deviations from the tripod design have been also studied~\cite{Trikha-Bhatia}. 

Many a time one is dealing with aspherical surfaces (non-spherical surfaces).  
For instance one may be dealing with cylindrical lenses or paraboloidal reflectors~\cite{Ghatak}.  
The study of elliptic and hyperbolic mirrors dates back to the time of Greeks~\cite{Hecht}
and the medieval Arabs.  A comprehensive account of the {\em Medieval Arab contributions to optics} 
can be found in~\cite{Rashed-1}-\cite{Khan-Refraction-OPN}.  
The beginning of the analogy between geometrical optics and mechanics, usually attributed to 
Descartes, can be traced to Abu Ali al-Hasan ibn al-Haytham (965-1039, known as Alhacen/Alhazen, 
the Latin transliteration of his first name al-Hasan, see \cite{Ambrosini}-\cite{Khan-FW-Advances} for details). 
 
The occurrence of the aspherical surfaces in the optics workplace necessitate devices beyond the common spherometer.  
One of the earliest modifications can be traced back to 1924 (see ~\cite{Rowell} for the details).  
A prime example of such a modification is the device, cylindrometer which is also known as 
the {\em Cylindro-Spherometer} and {\em Sphero-Cylindrometer}.  The cylindrometer can additionally 
measure the radius of curvature of a right-circular 
cylinder~\cite{Cylindrometer-Singh}-\cite{Khan-cylindrometer-iapt}.  
It is to be noted that the cylindrometer can not distinguish between spherical and cylindrical 
surfaces.  Its usage requires one to assume that the surface under study is either spherical or 
cylindrical!  It is further to be noted that the design of both the spherometer and the cylindrometer 
are based on a geometric relation unique to circles and spheres, known since the times of 
Euclid~\cite{Hecht}.  
 
In this article, we present an alternate approach using coordinate geometry in place of the 
Euclidean geometry.  This provides us the familiar result for the spherometer.  
The coordinate geometric approach to the spherometer was developed recently~\cite{Khan-cg-iapt}. 
This approach, using the powerful techniques of coordinate geometry is suitable to 
a generalization of the traditional spherometer to devices, which can characterize 
aspherical surfaces~\cite{Khan-quadratic-surfaces}. 
Section-2 has a review of the traditional derivation of the spherometer formula. 
Section-3 has a comprehensive account of the coordinate geometric approach to the 
spherometer. 
In section-4 the coordinate geometric approach is extended to the cylindrometer.  
In the Appendices, we shall describe the variants and modifications of the common spherometer. 
We shall also consider other geometries such as regular polygons and the ring geometry. 
Appendix-A covers the {\em ball-spherometer}, which safeguards the surfaces under study from 
the sharp tips of the spherometer legs.  
Appendix-B describes the ring-spherometer.  
Both of these appendices contain images of the instruments fabricated by the author.  
Appendix-C describes the cylindrometer which can be used to measure 
the radius of curvature of both spherical and cylindrical surfaces.  
This appendix also covers the {\em ball-cylindrometer} fabricated by the author.  
Appendix-D has a note on the aspherical surfaces.  
Appendix-E is dedicated to the class of surfaces known as the quadratic surfaces. 
Appendix-F contains the abstract of the {\em quadricmeter} invented by the author. 
This device can completely characterize the quadratic surfaces.  
The spherometer, cylindrometer and the quadricmeter are contact devices.  
Appendix-G is dedicated to optical techniques used to characterize surfaces. 
Appendix-H has a short note on the radius of curvature with a mathematical perspective.

\section{Spherometer: the Traditional Approach} 
Since the beginning the spherometers consist of a tripod framework supported on 
three fixed legs of equal lengths.  Ideally the tips of the three legs form an 
equilateral triangle.  The deviations from the ideal equilateral triangular geometry 
shall be discussed at the end of this section.  A precisely cut micrometer-screw is 
made to pass through a nut fixed at the centroid of the equilateral triangle.  
The movable micrometer-screw and the three fixed legs should all be parallel to 
each other.  
The spherometer has two scales.  One is the common millimeter scale, attached 
vertically to the tripod frame such that it is parallel to the axis o the 
micrometer-screw.  The second scale is a large circular disc (with typically a 
hundred divisions), which is attached to the top of the micrometer-screw.  
The two scales together, working on the principle of the {\em screw gauge} provide 
an accurate measurement of the relative height of the tip of the micrometer-screw 
with respect to the plane containing the tips of the three fixed legs.  This height 
of the micrometer-screw is known as {\em sagitta}.  Accuracy requires the tips to be 
as sharp as possible. 

Let the spherometer be placed on any spherical surface of radius of curvature $R$.  
The great circle of radius $R$ touches the tip of the micrometer-screw, which has a 
relative height $h$ with respect to the plane containing the (equilateral) triangle 
formed by the tips of the three fixed legs.  Let $L$ be the side of the equilateral 
triangle.  

\begin{figure}[ht]
\begin{center}
\includegraphics[width=8cm]{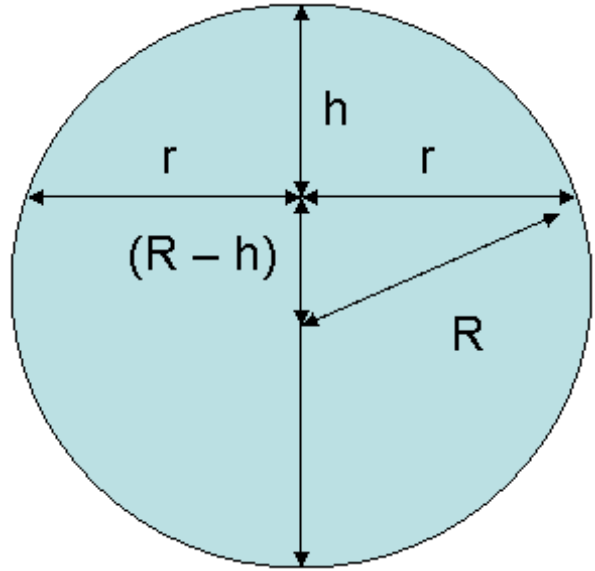} 
\caption{The Geometry of the Spherometer.}
\label{figure-spherometer-euclidean-geometry}
\end{center}
\end{figure}

\noindent
In Figure-\ref{figure-spherometer-euclidean-geometry}, we apply the Pythagorean 
theorem to the triangle and obtain the relation 
\begin{eqnarray}
r^2 = R^2 - (R - h)^2 = h(2R - h)\,. 
\label{spherometer-central}
\end{eqnarray}
This relation is central to the derivation of the spherometer formula.  
In fact such a relation exists for any two chords (which need not be perpendicular, 
as shown in the figure-\ref{figure-spherometer-euclidean-geometry}), 
making circle unique with respect to the other conic sections.  
Equation~(\ref{spherometer-central}) leads to the following equivalent relations 
\begin{eqnarray}
R(h) & = & \frac{r^2}{2 h} + \frac{h}{2}\,, \nonumber \\
h(r) & = & R - \sqrt{R^2 - r^2} = \frac{r^2}{R + \sqrt{R^2 - r^2}}\,.
\label{spherometer-basic}
\end{eqnarray}
The radius, $r$ of the circumcircle and the length, $L$ of the equilateral triangle 
obey the relation $r = L/\sqrt{3}$.  So, the radius of the sphere is given by   
\begin{eqnarray}
R(h) = \frac{L^2}{6 h} + \frac{h}{2}\,, 
\label{spherometer-sphere}
\end{eqnarray}
where $h$ is the only measurable variable and $L$ is a device constant. 
For a given instrument, the smallest measurable radius of curvature 
is $R_{\rm min} = R (h = L/{\sqrt{3}}) = r = L/{\sqrt{3}}$, 
which is at $h = r = L/{\sqrt{3}}$.  
The largest radius of curvature is constrained by the least count
of the micrometer-screw (typically $0.01mm$ or better).  The side $L$ of the spherometer 
is several centimeters or more depending on the size of spherical surface to be studied.  
The size of a spherometer depends on the application.  To measure just the sagitta at 
the centre of a mirror, one requires a spherometer with a base almost as big as the 
mirror itself.  Using a smaller base one can move around the spherometer to detect local 
high and low spots produced during the fabrication.
 
Since the spherometer is essentially a type of micrometer, it can be employed for purposes 
other than measuring the curvature of a spherical surface.  For example, it can be used to 
measure the thickness of a thin plate.  To do so, the instrument is placed on a level 
plane surface and the screw turned until the point just touches it.  Scales are read; 
screw is raised; the thin plate slipped under it; and the process is repeated. 
The difference between the two readings gives the required thickness.
 
During constructions there can be deviations from the ideal equilateral triangle geometry.  
When the lengths of the sides of the triangle are $L_1$, $L_2$ and $L_3$ respectively, 
the radius $r$ of the circumcircle is given by 
\begin{eqnarray}
r = \frac{L_1 L_2 L_3}{4 \sqrt{s (s - L_1 ) (s - L_2 ) (s - L_3 )}}\,,
\label{deviations}
\end{eqnarray}
where $s = {(L_1 + L_2 + L_3)}/2$ is the semi-perimeter of the triangle.  
Using the relation~(\ref{deviations}), we obtain 
\begin{eqnarray}
R = \frac{L_1^2 L_2^2 L_3^2}{32 h s (s - L_1 ) (s - L_2 ) (s - L_3 )} + \frac{h}{2}\,. 
\label{r-deviations}
\end{eqnarray}
The formula in~(\ref{r-deviations}) generalizes the formula for the equilateral triangle 
in~(\ref{spherometer-sphere}) and is due to Trikha and Bhatia (see \cite{Trikha-Bhatia} 
for the details). 
 
The derivation, though elegant explicitly depends on two facts: firstly that the 
plane-sections of a sphere are always circles and secondly the geometric 
relation in~(\ref{spherometer-central}).  It is well known that the plane-sections of 
aspherical surfaces are in general not circles, but some other curves such as conic 
sections (ellipse, parabola and hyperbola).  Moreover the corresponding geometric 
relations do not exist for conic sections.  Analogous relations do exist for ellipse, 
parabola and hyperbola but work only at a single point, the vertex with $h$ aligned to 
the axis of each conic respectively.  Thus any attempt to design a spherometer-type 
device, based on the geometric relations would work only at one point, whose location 
would of course not be known a priory.  So, it is natural to look beyond the geometric 
relations.  The coordinate geometry provides a very natural choice~\cite{Khan-cg-iapt}.

\section{Spherometer: A Coordinate Geometric Representation}
\begin{figure}[ht]
\begin{center}
\includegraphics[width=8cm]{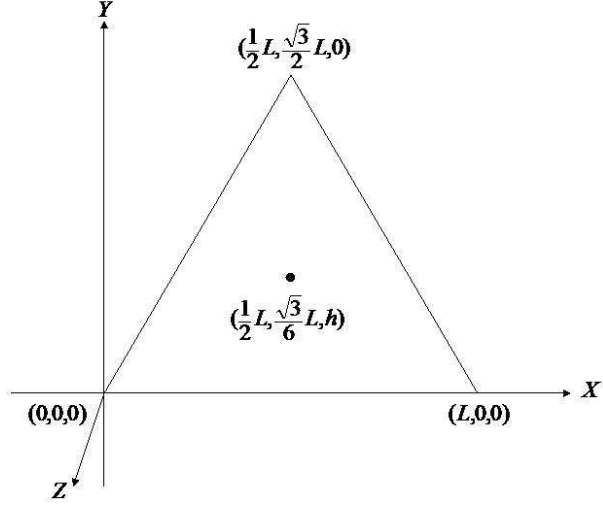} 
\caption{A Coordinate Representation of the Spherometer.}
\label{figure-spherometer-coordinate-geometry}
\end{center}
\end{figure}

In the framework of the coordinate geometry, any geometric figure is specified by a certain 
number of points on it.  For instance, a sphere is uniquely determined by specifying four 
points on it.  The four points are of course not coplanar~\cite{Shanti-Narayan}.  
The situation in the spherometer is precisely the same!  
The three points are the tips of the fixed legs forming the equilateral triangle and the 
fourth point is the tip of the movable micrometer-screw.  
This forms the basis of the coordinate geometric approach to the
spherometer.  The coordinate geometric representation of a sphere is not unique.  A possible 
representation is as follows.  We choose the plane of the equilateral triangle to lie completely 
in the $X$-$Y$~plane, without any loss of generality.  Then the tip of the micrometer-screw 
moves parallel to the $Z$-axis.  Figure-\ref{figure-spherometer-coordinate-geometry} has the 
details of the coordinate geometric representation. 

The choice of our coordinate geometric representation is such that when the spherometer 
is placed on any plane surface, the $z$-coordinates of all the four points are identically 
zero.  This is called as the {\em ground state} of the spherometer.  This representation 
is such that, when the spherometer is placed on any spherical surface, only the $z$-coordinate 
of the micrometer-screw acquires the non-zero value of $h$; the remaining three corresponding to
the three fixed legs remain unchanged from the value zero. 

The most general equation of a sphere of radius $R$ and centre at the point $(\alpha, \beta, \gamma )$ is 
\begin{eqnarray}
(x - \alpha )^2 + (y - \beta )^2 + (z - \gamma )^2 = R^2\,. 
\label{sphere-equation}
\end{eqnarray}
The aforementioned equation has four unknown constants (including the radius, $R$), which can be 
uniquely determined by specifying four points on the sphere.  Let us recall that the four points 
can not be coplanar.
In our choice of the coordinate representation, using the coordinates of the tips of the three 
fixed legs, we readily obtain, $\alpha = L/2$, $\beta = (\sqrt{3}/6)L$, $\gamma^2 = R^2 - L^2/3$ 
and $\alpha^2 + \beta^2 + \gamma^2 = R^2$.  
The coordinates of the movable tip of the micrometer-screw lead decisively 
to $\gamma = {L^2}/{6h} - {h/2}$.  Using these values of the four constants, the radius $R$ of 
the spherical surface is obtained after straightforward substitution 
\begin{eqnarray}
R = \sqrt{\alpha^2 + \beta^2 + \gamma^2} = \pm \left(\frac{L^2}{6 h} + \frac{h}{2} \right)\,.
\label{spherometer-sphere-cg}
\end{eqnarray}
The {\em positive} and {\em negative} signs are for the {\em convex} and {\em concave} 
surfaces respectively.  
 
The technique of the coordinate geometry has enabled us to understand the working of the 
spherometer in an entirely different manner.  The traditional approach is based on a 
geometric relation unique to circles and spheres.  The newly developed coordinate geometric 
approach is independent of any such explicit geometric relations.  The choice of the 
coordinate representation ensures that the sphere is completely characterized by a single 
parameter, that is the height of the movable micrometer-screw.  There are several advantages 
of using the coordinate geometric approach.  It is capable of handling any changes, which may 
arise during the fabrication of the spherometer or during the measurements.  Such changes 
are not uncommon, particularly when the spherometer is large, then it can deviate from its 
equilateral triangle geometry.  In such a situation, it would suffice to revise the ground state of 
the spherometer, even at the final stages while the measurements are being made.  
The greater advantage of using the coordinate geometric approach is that it provides a better 
understanding of the spherometer design and further paves the way to modify and extend the 
tripod spherometer to devices which can handle surfaces which need not be spherical. 
Let us recall that the aspherical surfaces do not have the required geometric relations 
available for circles and spheres. 
 
The equation of the sphere in terms of the four points lying on it can be expressed in the form of the 
following determinant equation (see \cite{Shanti-Narayan}-\cite{Weisstein-Mathworld} for details). 
\begin{eqnarray}
{\left|
\begin{array}{ccccc}
x^2 + y^2 + z^2 & x & y & z & 1 \\ 
x_1^2 + y_1^2 + z_1^2 & x_1 & y_1 & z_1 & 1 \\
x_2^2 + y_2^2 + z_2^2 & x_2 & y_2 & z_2 & 1 \\
x_3^2 + y_3^2 + z_3^2 & x_3 & y_3 & z_3 & 1 \\
x_4^2 + y_4^2 + z_4^2 & x_4 & y_4 & z_4 & 1 
\end{array}
\right|} = 0\,.
\label{sphere-determinant}
\end{eqnarray}
For the common spherometer with an equilateral triangle base, and our choice of the coordinate 
representation in Figure-\ref{figure-spherometer-coordinate-geometry}, the determinant 
equation in~(\ref{sphere-determinant}) translates to  
\begin{eqnarray}
{\left|
\begin{array}{ccccc}
x^2 + y^2 + z^2 & x & y & z & 1 \\ 
0 & 0 & 0 & 0 & 1 \\
L^2 & L & 0 & 0 & 1 \\
L^2 & \frac{1}{2} L & \frac{\sqrt{3}}{2} L & 0 & 1 \\
\frac{1}{3} L^2 + h^2 & \frac{1}{2} L & \frac{\sqrt{3}}{6} L & h & 1 
\end{array}
\right|} = 0\,.
\label{spherometer-determinant}
\end{eqnarray}
The above equation can be solved to describe the sphere completely.  
One may use MS EXCEL~\cite{EXCEL}-\cite{EXCEL-IAPT} or a versatile symbolic package 
such as the MATHEMATICA incorporating the graphic environment~\cite{MATHEMATICA, MATHEMATICA-Boccara}.

\section{Cylindrometer: A Coordinate Geometric Representation}
The traditional approach to the cylindrometer is described in detail in Appendix-C. 
The coordinate geometric approach to the cylindrometer is very similar to the one for
the spherometer.  
 
\begin{figure}[ht]
\begin{center}
\includegraphics[width=8cm]{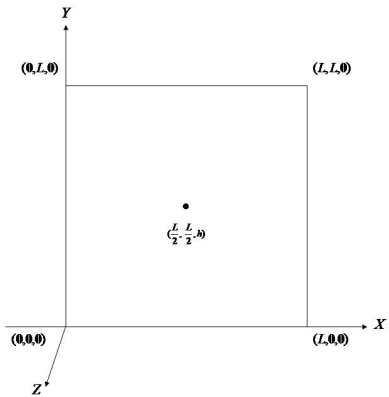} 
\caption{A Coordinate Representation of the Cylindrometer.}
\label{figure-cylindrometer-coordinate-geometry}
\end{center}
\end{figure}

A right-circular cylinder is uniquely determined by specifying five points 
on it~\cite{Shanti-Narayan}.  
This is precisely the situation in a cylindrometer: the four points are the fixed tips 
of the square-base and the fifth point is the movable tip of the micrometer-screw.  
The coordinate geometric representation of a cylinder is not unique and we have chosen 
the following representation, which is the simplest.   
The tips of the four fixed legs lying on the corners of the square of side $L$ 
can be chosen to lie completely in the $X$-$Y$~plane, without any loss of generality.
The tip of the micrometer-screw (lying at the centre of the square) moves parallel to 
the $Z$-axis.  
Figure-\ref{figure-cylindrometer-coordinate-geometry} has the required details.    
 
In this choice of representation, when the cylindrometer is placed on a plane surface, 
the $z$-coordinates of all the five points are identically zero.  We call this as 
the {\em ground state} of the cylindrometer.  When the cylindrometer is placed on a 
cylindrical surface or a spherical surface, only the $z$-coordinate of the 
micrometer-screw changes to $h$ and the remaining four points remain unchanged from 
the value zero. 
 
The general equation of a right-circular cylinder of radius $R$ and whose axis is the 
line ${(x - \alpha)}/l = {(y - \beta)}/m = {(z - \gamma)}/n$ is 
\begin{eqnarray}
(x - \alpha )^2 + (y - \beta )^2 + (z - \gamma )^2 
- \frac{[l(x - \alpha) + m (y - \beta) + n (z - \gamma)]^2}{l^2 + m^2 + n^2 } = R^2\,. 
\label{cylinder-equation}
\end{eqnarray}
It is to be noted that the square base ensures that one pair of opposite sides of the 
square base are parallel to the axis of the cylinder and the other pair is perpendicular.  
This aspect helps us in simplifying the general equation of the cylinder.  
Without any loss of generality let us assume that the axis of the cylinder is parallel to 
the $X$-axis.  Then, we have $l = 1$ and $m = 0 = n$.  This reduces the general equation 
to $(y - \beta )^2 + (z - \gamma )^2 = R^2$.  We handle this equation in a manner very 
similar to the spherometer and obtain 
\begin{eqnarray}
R_{\rm cylinder} = \pm \left(\frac{L^2}{8 h} + \frac{h}{2} \right)\,.
\label{cylindrometer-cylinder-cg}
\end{eqnarray}

Now, we place the cylindrometer on a sphere.  Incorporating the coordinates of the tips 
of the four fixed legs in the equation of the sphere in~(\ref{sphere-equation}) lead 
to $\alpha = \beta = L/2$ and $\gamma^2 = R^2 - {L^2}/{2}$. 
The coordinates of the movable tip of the micrometer-screw lead decisively 
to $\gamma = h \mp R$.  Eliminating $\gamma$, we readily obtain 
\begin{eqnarray}
R_{\rm sphere} = \pm \left(\frac{L^2}{4 h} + \frac{h}{2} \right)\,.
\label{cylindrometer-sphere-cg}
\end{eqnarray}
The {\em positive} and {\em negative} signs in~(\ref{cylindrometer-cylinder-cg}) and 
(\ref{cylindrometer-sphere-cg}) are for the convex and concave surfaces respectively.

Thus, we reproduce the familiar result using coordinate geometry.  
The choice of the coordinate representation ensures that the equation of the cylinder 
is dependent only on one measurable parameter, namely the height $h$ of the screw.  
The sphere and the right-circular cylinder are the only geometries, which can be handled 
using the cylindrometer.  Even a spheroid (ellipsoid of revolution) and the right-circular 
cone can not be handled using a cylindrometer.  This dictates the need to have newer 
instruments which can handle the other geometries.

\section{Concluding Remarks}
We have reviewed the common spherometer and its modified avatars such as the 
ring-spherometer and the ball-spherometer.  We have also reviewed the lesser known 
cylindrometer and seen how it can be used to measure the radii of curvature of 
both right-circular cylindrical surfaces and spherical 
surfaces~\cite{Khan-cylindrometer-aapt, Khan-cylindrometer-iapt}.  
The idea of the {\em ball-spherometer} has been extended to the cylindrometer 
resulting in a {\em ball-cylindrometer} (Appendices A and C have the details along 
with the images). 
It is essential to look beyond and generalize the common spherometer and the 
cylindrometer to devices, which can enable the study of a wider range of surfaces 
occurring in science and engineering~\cite{Khan-quadratic-surfaces}.  
Coordinate geometry provides a natural framework for creation of such 
devices~\cite{Khan-cg-iapt}.
 
In coordinate geometry, a given figure is uniquely determined by specifying a certain 
number of points on it.  This number varies from figure to figure.  For instance, a 
sphere is completely characterized by its radius and the four points lying on it are 
sufficient to fix its radius.  This demands the spherometer to be a four-point device.  
Hence its design has four points: the three points are the tips of the three fixed legs 
and the fourth point is the tip of the movable micrometer-screw.  The choice of the 
coordinate representation ensures that there is only one variable parameter, $h$ 
corresponding to the radius $R$ of the sphere.  A right-circular cylinder is 
characterized by its radius (the length of the cylinder is not under discussion in the 
present context), which can be fixed by five points on it.  Cylindrical symmetry permits 
points to be coplanar. 
Hence, the cylindrometer is a five-point device.  A judicious choice of the 
coordinate representation ensures that, we are again able to express the radius $R$ of 
the cylindrical surface in terms of a single device variable, which is again the height 
of the movable tip of the micrometer-screw of the cylindrometer. 
A quadratic surface is described by the general second-order equation, which has 
nine independent constants~\cite{Khan-quadratic-surfaces, Weisstein-Handbook}. 
The nine points (no four points being coplanar) uniquely determine the quadratic surface.  
So, we require a nine point device to identify and completely characterize any quadratic surface.  
The four-point coordinate representation of a spherometer 
and the five-point coordinate representation of a cylindrometer, has been generalized to 
a nine-point device, the quadricmeter, which generates the equation of the quadratic 
surfaces in terms of measurable laboratory parameters~\cite{quadricmeter, generalized-spherometer}.  
The quadricmeter is the instrument devised to identify, distinguish and measure the various 
parameters (including the axis, foci, latera recta, directrix) completely characterizing the 
important class of surfaces known as the quadratic surfaces.  
Quadratic surfaces (also known as quadrics and conicoids) include a wide range of 
commonly encountered surfaces including, cone, cylinder, ellipsoid, elliptic cone, 
elliptic cylinder, elliptic hyperboloid, elliptic paraboloid, hyperbolic cylinder, hyperbolic 
paraboloid, paraboloid, sphere, and spheroid.  
The complete characterization is done using the standard techniques of coordinate geometry.  
With a conventional spherometer it is possible only to measure the radii of spherical surfaces.  
Cylindrometer can measure the radii of curvature of a cylindrical surfaces in addition to the 
spherical surfaces.  In both the spherometer and the cylindrometer one has to necessarily assume 
the surface to be either spherical or cylindrical respectively.  In the case 
of the quadricmeter, there are no such assumptions~\cite{quadricmeter, generalized-spherometer} 
as it can distinguish one quadratic surface from the other.  
The nomenclature {\em quadricmeter} originates from the word {\em quadrics} used for quadratic surfaces 
and was preferred over {\em conicoidmeter}.

\setcounter{section}{0}
 
\section*{}
\addcontentsline{toc}{section}
{Appendix A. \\
Ball-Spherometer}

\renewcommand{\theequation}{A.{\arabic{equation}}}
\setcounter{equation}{0}
 
\begin{center}
 
{\Large\bf
Appendix A. \\
Ball-Spherometer
} \\

\end{center}
 
In order to ensure higher accuracy, the spherometer is required to have sharper legs.  
This is sure to harm the surfaces under study.  This is more true, when the surfaces 
are large (as in the telescope mirrors) and the required spherometers are bound to be 
heavy.  One of the most widely used methods to handle this situation is to replace 
the sharp tips of the three legs of the spherometer with balls of radius $r_0$.  
The resulting spherometer is called a {\em ball-spherometer} and the basic relation 
in~(\ref{spherometer-basic}) for the radius of curvature is modified to  
\begin{eqnarray}
R(h) = \frac{r^2}{2 h} + \frac{h}{2} \pm r_0\,,
\label{ball-spherometer}
\end{eqnarray}
where the positive and negative signs are for the convex and concave surfaces 
respectively~\cite{Karow}.  The idea of replacing the sharper tips with balls 
has been extended to the cylindrometer.  The author of this article strongly advocates 
to use the {\em ball-cylindrometer} resulting from this replacement.  

\begin{figure}[ht]
\begin{center}
\includegraphics[width=8cm]{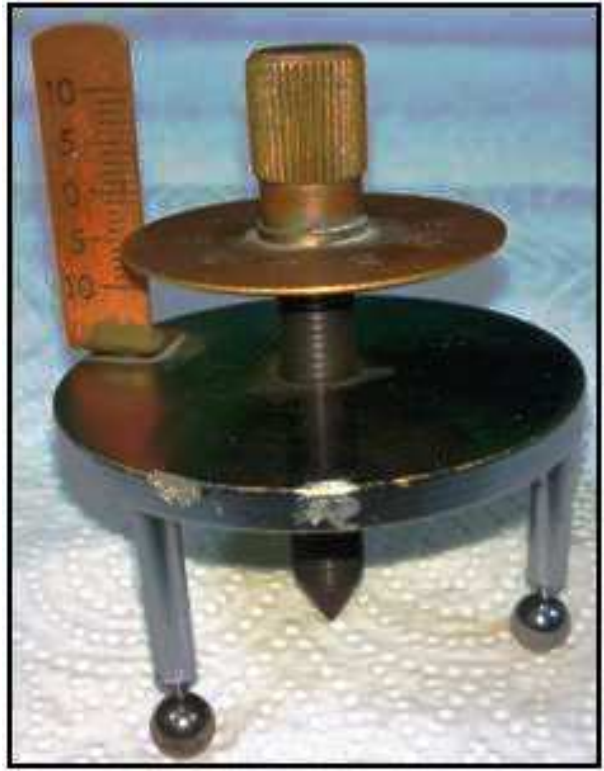} 
\caption{The Ball-Spherometer fabricated by the author.}
\label{figure-ball-spherometer}
\end{center}
\end{figure}

\setcounter{section}{0}
 
\section*{}
\addcontentsline{toc}{section}
{Appendix B. \\
Ring-Spherometer}

\renewcommand{\theequation}{B.{\arabic{equation}}}
\setcounter{equation}{0}
 
\begin{center}
 
{\Large\bf
Appendix B. \\
Ring-Spherometer
} \\

\end{center}
 
Large mirrors arise in the telescopes requiring larger spherometers, which are bound to
be heavy.  A possible solution is to replace the tripod legs with a continuous ring.  
The resulting device is called a {\em ring-spherometer}~\cite{Sirohi-Ring-Spherometer}.  
If $r$ is the radius of the ring, then the radius, $R$ of the spherical surface is  
\begin{eqnarray}
R(h) = \frac{r^2}{2 h} + \frac{h}{2}\,.
\label{ring-spherometer}
\end{eqnarray}
Ring-spherometers provide an average curvature of the surface.  The local structure of 
the same surface is obtained by using the regular spherometers of smaller sizes.  In 
order to ensure accuracy the ring need to be fairly sharp.  A finer calculation requires 
us to use the internal and external radii of the ring for convex and concave surfaces 
respectively.

\begin{figure}[ht]
\begin{center}
\includegraphics[width=8cm]{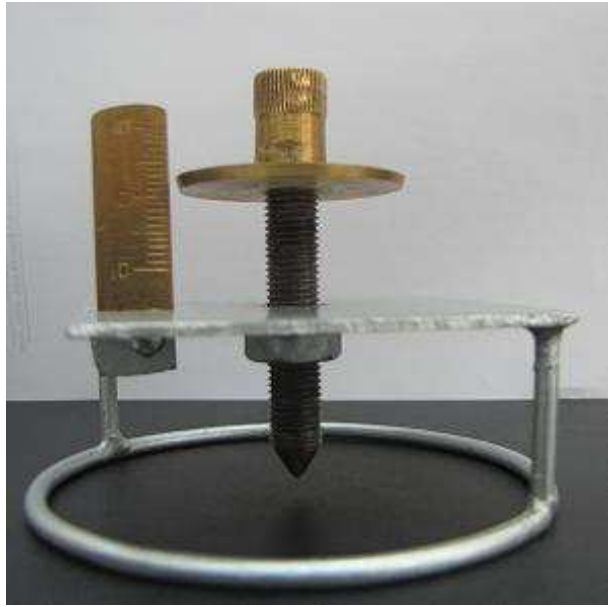} 
\caption{The Ring-Spherometer fabricated by the author.}
\label{figure-ring-spherometer}
\end{center}
\end{figure}

In principle, one can use any regular polygon inscribed in the circle of radius, $r$.  
The side $L$ of a regular $n$-sided polygon is related to the radius $r$ of the 
circumcircle by the relation $r = L/{2 \sin (\pi/n)}$.  This leads to 
\begin{eqnarray}
R(h) = \frac{L^2}{8 h \sin^2 (\pi/n)} + \frac{h}{2}\,.
\label{polygon-spherometer}
\end{eqnarray}

\setcounter{section}{0}
 
\section*{}
\addcontentsline{toc}{section}
{Appendix C. \\
Cylindrometer}

\renewcommand{\theequation}{C.{\arabic{equation}}}
\setcounter{equation}{0}
 
\begin{center}
 
{\Large\bf
Appendix C. \\
Cylindrometer
} \\

\end{center}

The cylindrometer is an old idea but seldom mentioned in the introductory laboratory courses.  
Reference to it can be traced back to an article published in 1924; but the design presented 
in it lacked accuracy~\cite{Rowell}.  
An ideal cylindrometer has a square framework supported on four fixed legs of equal lengths.  
The tips of the four legs lie on the corners of the square.  An accurately cut micrometer-screw, 
passes through the nut fixed at the centroid of the square (point where the two diagonals of 
the square intersect).  The screw is parallel to the four fixed legs.  A large circular disc 
with typically a hundred divisions is attached to the top of the screw.  
A small millimetre scale is vertically attached to the square framework (parallel to the axis 
of the screw).  The two scales together (working on the principle of the screw gauge) provide 
the relative height of the screw (known as {\em sagitta}) with respect to the tips of the four 
fixed legs.  The radius of curvature, $R$ of the cylindrical (and spherical) surface  
is related to the side of the square $L$ and the height of the micrometer-screw $h$ by 
geometric relations, which we shall derive.

\begin{figure}[ht]
\begin{center}
\includegraphics[width=8cm]{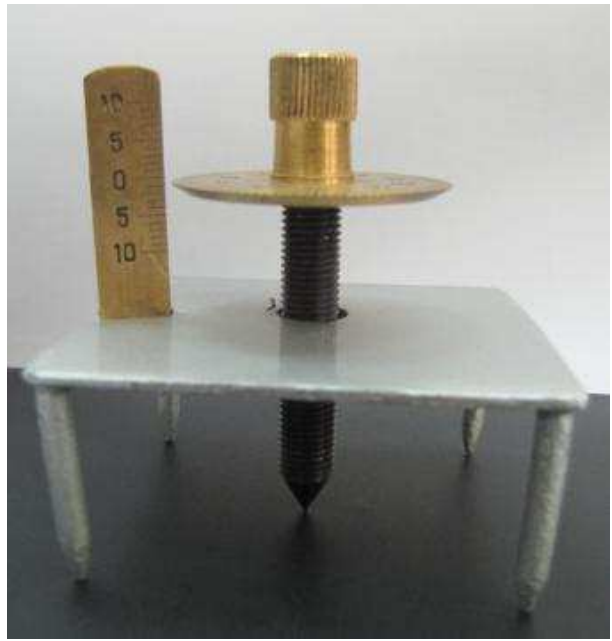} 
\caption{The Cylindrometer fabricated by the author.}
\label{figure-cylindrometer}
\end{center}
\end{figure}

The cylindrometer, when used on a spherical surface, the underlying geometry is a circle 
with the square inscribed in it; the great circle (whose radius is $R$) touches the tip of the 
screw of height $h$, measured relative to the plane containing the tips of the four fixed legs.  
When the cylindrometer is placed on a cylindrical surface, the two opposite sides of the square 
base get aligned parallel to the axis of the cylinder; consequently the tips of two legs lie 
on the circle defining the cylinder and the height $h$ of the screw is measured relative to 
the plane containing the tips of these four legs.

For the cylinder, $r$ is related to the side, $L$ of the square by the relation, $r = L/2$, 
and we obtain 
\begin{eqnarray}
R_{\rm cylinder} = \frac{L^2}{8 h} + \frac{h}{2}\,.
\label{cylindrometer-cylinder}
\end{eqnarray}
For the sphere, $r = L/{\sqrt{2}}$, leading to 
\begin{eqnarray}
R_{\rm sphere} = \frac{L^2}{4 h} + \frac{h}{2}\,.
\label{cylindrometer-sphere}
\end{eqnarray}
The above relations are structurally similar to the formula in Equation~(\ref{spherometer-sphere}) 
for a common spherometer with the tripod base.  
For a given instrument, the smallest measurable radius of curvature for a cylinder and sphere 
are $R_{\rm cylinder}^{\rm min} = L/2$ and $R_{\rm sphere}^{\rm min} = L/{\sqrt{2}}$ respectively.  
The largest radius of curvature is constrained by the least count of the micrometer-screw 
($0.01mm$ or better).  The side $L$ of the cylindrometer is several centimetres or more depending 
on the size of surface to be studied.  The size of a cylindrometer depends on the application.  
To measure just the sagitta at the centre of a mirror, one requires a cylindrometer with a base 
almost as big as the mirror itself.  Using a smaller base one can move around the cylindrometer 
to detect local high and low spots produced during the fabrication.

The design of the cylindrometer with a square base was patented in 1972 by Gur Iqbal Singh Hunjan, 
with the title, {\em Cylindro-Spherometer} (see \cite{Cylindrometer-Singh} for complete specifications).
The other name of the device is {\em Sphero-Cylindrometer}.   
It is very surprising that the cylindrometer is neither mentioned in the introductory books and nor 
it is available from the scientific instruments companies.  
Cylindrometers fabricated in-house are used in many laboratories world-wide~\cite{Cylindrometer-In-House}.

\begin{figure}[ht]
\begin{center}
\includegraphics[width=8cm]{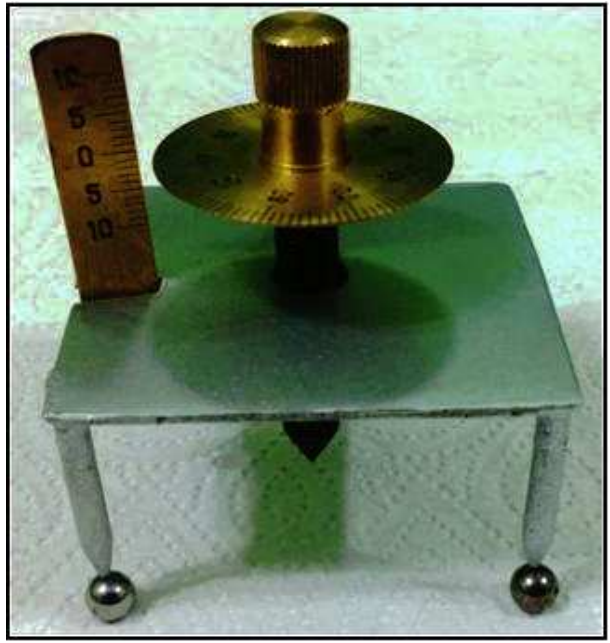} 
\caption{The Ball-Cylindrometer fabricated by the author.}
\label{figure-ball-cylindrometer}
\end{center}
\end{figure}

The cylindrometer too has the issue of the sharper legs damaging the surfaces under study.  
It is straightforward to replace the tips of the sharp legs with balls of radius $r_0$.  
The resulting device is called as the {\em ball-cylindrometer} 
(see Figure-\ref{figure-ball-cylindrometer}) in analogy with the {\em ball-spherometer}. 
It works on both cylindrical and spherical surfaces.  
The corresponding radii of curvature are given by 
\begin{eqnarray}
R_{\rm cylinder} & = & \frac{L^2}{8 h} + \frac{h}{2} \pm r_0\,, \nonumber \\
R_{\rm sphere} & = & \frac{L^2}{4 h} + \frac{h}{2} \pm r_0\,,
\label{ball-cylindrometer-radius}
\end{eqnarray}
where the positive and negative signs are for the convex and concave surfaces respectively.

\setcounter{section}{0}
 
\section*{}
\addcontentsline{toc}{section}
{Appendix D. \\
Aspherical Surfaces}

\renewcommand{\theequation}{D.{\arabic{equation}}}
\setcounter{equation}{0}
 
\begin{center}
 
{\Large\bf
Appendix D. \\
Aspherical Surfaces
} \\

\end{center}
 
One often needs to quantify aspherical surfaces, such as the parabolic mirrors.
The traditional spherometers can be used to a certain extent for this purpose.  
Many aspheric surfaces can be approximated as conic sections of revolution.  
Conic sections are generally easier to test than a general asphere, 
because there are geometric null tests for conics~\cite{Stoltzmann}.  
Incorporating the conic constant $K = - e^2$, where $e$ is the eccentricity  
of the conic section, the basic equation~(\ref{spherometer-basic}) modifies to 
\begin{eqnarray}
z(r) = \frac{r^2}{R + \sqrt{R^2 - (K + 1) r^2}}\,.
\label{spherometer-aspheric}
\end{eqnarray}
In this equation $z(r) = h(r)$ is the surface height, $r = \sqrt{(x^2 + y^2)}$
is the radial position and $R$ is the radius of curvature. 
The types of conic surfaces, determined by the conic constant $K = - e^2$, are 
as follows 

\medskip

\begin{center}
\begin{tabular}{|l|l|}
\hline 
$K < - 1$ & Hyperboloid \\ 
\hline
$K = - 1$ & Paraboloid \\
\hline
$- 1 < K < 0$ & Prolate Ellipsoid (rotated about its major axis) \\
\hline
$K = 0$ & Sphere \\ 
\hline
$K > 0$ & Oblate Ellipsoid (rotated about its minor axis) \\ 
\hline
\end{tabular} \\
{\bf Table-1:} Types of Conic Surfaces \\
\end{center}
 
\noindent
A series expansion of equation~(\ref{spherometer-aspheric}) can be used to calculate 
the aspheric departure~\cite{Anderson-Burge}.  
If the location of the vertex is known, then a pair of readings using two 
ring-spherometers of different sizes, close to the vertex can be useful in assessing 
such surfaces.  However, such a procedure would require lot of skill.

\setcounter{section}{0}
 
\section*{}
\addcontentsline{toc}{section}
{Appendix E. \\
Quadratic Surfaces}

\renewcommand{\theequation}{E.{\arabic{equation}}}
\setcounter{equation}{0}

\renewcommand{\thesection}{E.{\arabic{section}}} 
\setcounter{section}{1}

\begin{center}
 
{\Large\bf
Appendix E. \\
Quadratic Surfaces
} \\

\end{center}
 
Many a time one is dealing with aspherical surfaces (non-spherical surfaces).  
For instance one may be dealing with cylindrical lenses or paraboloidal 
reflectors~\cite{Ghatak}. 
The study of elliptic and hyperbolic mirrors dates back to the time of 
Greeks~\cite{Hecht} and the medieval Arabs~\cite{Rashed-1}-\cite{Khan-Refraction-OPN}.
In telescopes and other optical instruments, the mirrors and lenses generally 
have a shape coming from quadratic surfaces~\cite{Malacara, Karow}.  
One also finds them in aerodynamic modeling.  Such situations, involving 
aspherical surfaces are very common across almost all scientific disciplines.  
Quadratic surfaces (also known as quadrics and conicoids) include a wide 
range of commonly encountered surfaces including, cone, cylinder, ellipsoid, 
elliptic cone, elliptic cylinder, elliptic hyperboloid, elliptic paraboloid, 
hyperbolic cylinder, hyperbolic paraboloid, paraboloid, sphere, and 
spheroid~\cite{Weisstein-Handbook}.  From the aforementioned list it is 
evident that one comes across the quadratic surfaces in various areas 
of science and engineering.  Quadratic surfaces have been a basic primitive 
in computer graphics for a long time and their properties and techniques for 
visualization are very well researched.  Quadratic surfaces are important 
in geometric modeling.  Modeling of three dimensional objects with quadratic 
surface patches is very effective.  Complicated surfaces can be approximated as 
patches of quadratic surfaces.  In this appendix, we shall describe the algebraic 
classification of the quadratic surfaces into the seventeen standard-form types 
along with their corresponding equations.  We shall also state the conditions under 
which a given quadric is a {\em quadratic surface of revolution}.

\subsection{Classification of Quadratic Surfaces}
A quadratic surface is described by the general second-order equation 
\begin{eqnarray}
ax^2 + b y^2 + c z^2 + 2 f y z + 2 g z x + 2 h x y + 2 p x + 2 q y + 2 r z + d = 0\,. 
\label{quadratic-surface-equation}
\end{eqnarray}
Equation~(\ref{quadratic-surface-equation}) has ten constants of which one is redundant 
since the equation can be divided by any one of the constants leading to nine 
independent constants.  Plane sections of a quadratic surface are conic sections namely 
ellipse, parabola and hyperbola (see~\cite{Shanti-Narayan, Dresden} for details).  
Quadratic surfaces have been fully classified and there are 17 standard-form 
types~\cite{Weisstein-Handbook}.  These are enumerated in Table-1.  Higher order surfaces 
have been studied extensively.  Cubic surfaces are algebraic surface of order three and 
have been fully classified.  However, quartic surfaces and higher order surfaces have 
not been fully classified~\cite{Weisstein-Handbook, Weisstein-Mathworld}. 

In order to classify the quadratic surfaces, following the notation in~\cite{Weisstein-Handbook}, 
we define  
\begin{eqnarray}
e & = &
{\left[
\begin{array}{ccc}
a & h & g \\
h & b & f \\
g & f & c  
\end{array}
\right]} \nonumber \\
E & = & 
{\left[
\begin{array}{cccc}
a & h & g & p \\
h & b & f & q \\
g & f & c & r \\ 
p & q & r & d 
\end{array}
\right]} \nonumber \\  
\rho_3 & = & rank(e) \nonumber  \\ 
\rho_4 & = & rank(E)  \nonumber  \\ 
\Delta & = & det (E)\,. 
\label{quadratic-surface-notation}
\end{eqnarray}
Let $k_1$, $k_2$ and $k_3$ be the eigenvalues (characteristic roots) of the matrix $e$, 
that is the roots of the characteristic equation 
\begin{eqnarray}
{\left|
\begin{array}{ccc}
a - x & h & g \\
h & b - x & f  \\
g & f & c - x  
\end{array}
\right|} = 0\,.
\label{quadratic-surfaces-characteristic-equation}
\end{eqnarray}
We further define 
\begin{eqnarray}
k = 
\left\{
\begin{array}{ll}
1 & \mbox{if the signs of the nonzero $k$s are the same}, \\ 
0 & \mbox{otherwise}.
\end{array}
\right. 
\label{conditions-on-k}
\end{eqnarray}
Using the above defined quantities the seventeen quadrics and their equations in reduced form, 
are listed in Table-2.  It is to be noted that sphere is not separately listed as it is a 
special case of the ellipsoid ($a = b = c$).  Likewise, the right-circular cone (special case 
of the elliptic cone, $a = b$) and the right-circular cylinder (special case of the 
elliptic cylinder, $a = b$) are also not listed separately.  
Any quadratic surface described by the general equation~(\ref{quadratic-surface-equation}) 
can be transformed to one of the seventeen standard-form types using translations and 
rotations of the coordinate axes. 
 
\medskip

\begin{center}
\begin{tabular}{|c|l|l|c|c|c|c|}
\hline
{\bf S. No.} & {\bf Surafce} & {\bf Equations} & \multicolumn{4}{c|}{{\bf Conditions}} \\ 
\cline{4-7}
 & & & $\rho_3$ & $\rho_4$ & sign$(\Delta)$ & $k$ \\
\hline
1 & Coincident Planes & $x^2 = 0$ & 1 & 1 & & \\ 
\hline
2 & Ellipsoid (Imaginary) & $\frac{x^2}{a^2} + \frac{y^2}{b^2} + \frac{z^2}{c^2} = - 1$ & 3 & 4 & $+$ & 1 \\ 
\hline
3 & Ellipsoid (Real) & $\frac{x^2}{a^2} + \frac{y^2}{b^2} + \frac{z^2}{c^2} = 1$ & 3 & 4 & $-$ & 1 \\  
\hline
4 & Elliptic Cone (Imaginary) & $\frac{x^2}{a^2} + \frac{y^2}{b^2} + \frac{z^2}{c^2} = 0$ & 3 & 3 & & 1 \\
\hline
5 & Elliptic Cone (Real) & $\frac{x^2}{a^2} + \frac{y^2}{b^2} = z^2$ & 3 & 3 & & 0 \\
\hline
6 & Elliptic Cylinder (Imaginary) & $\frac{x^2}{a^2} + \frac{y^2}{b^2} = - 1$ & 2 & 3 & & 1 \\
\hline
7 & Elliptic Cylinder (Real) & $\frac{x^2}{a^2} + \frac{y^2}{b^2} = 1$ & 2 & 3 & & 1 \\
\hline
8 & Elliptic Paraboloid & $\frac{x^2}{a^2} + \frac{y^2}{b^2} = z$ & 2 & 4 & $-$ & 1 \\
\hline
9 & Hyperbolic Cylinder & $\frac{x^2}{a^2} - \frac{y^2}{b^2} = - 1$ & 2 & 3 & & 0 \\
\hline
10 & Hyperbolic Paraboloid & $\frac{x^2}{a^2} - \frac{y^2}{b^2} = - z$ & 2 & 4 & $+$ & 0 \\
\hline
11 & Hyperboloid of one sheet & $\frac{x^2}{a^2} + \frac{y^2}{b^2} - \frac{z^2}{c^2} = 1$ & 3 & 4 & $+$ & 0 \\ 
\hline
12 & Hyperboloid of two sheets & $\frac{x^2}{a^2} + \frac{y^2}{b^2} - \frac{z^2}{c^2} = - 1$ & 3 & 4 & $-$ & 0 \\ 
\hline
13 & Intersecting Planes (Imaginary) & $\frac{x^2}{a^2} + \frac{y^2}{b^2} = 0$ & 2 & 2 & & 1 \\ 
\hline
14 & Intersecting Planes (Real) & $\frac{x^2}{a^2} - \frac{y^2}{b^2} = 0$ & 2 & 2 & & 0 \\ 
\hline
15 & Parabolic Cylinder & $x^2 + 2 r z = 0$ & 1 & 3 & & \\ 
\hline
16 & Parallel Planes (Imaginary) & $x^2 = - a^2 $ & 1 & 2 & & \\ 
\hline
17 & Parallel Planes (Real) & $x^2 = a^2$ & 1 & 2 & & \\ 
\hline
\end{tabular} \\
\label{table-2-quadratic-surfaces}
{\bf Table-2:} Seventeen Standard-Form Types of Quadrics \\
\end{center}

In passing, we note that the classification of the quantum mechanical quadratic 
Hamiltonians into classes with a physically relevant representation has been done.  
In the context of optics such a classification has been done by enumerating the 
Hamiltonian orbits~\cite{Simon-1}-\cite{Khan-Wolf}. 
For a comprehensive account of the Hamiltonian Orbits, see the book in~\cite{Wolf}. 
 

\subsection{Quadratic Surfaces of Revolution}
Quadratic surfaces of revolution (cones, cylinders, ellipsoids, hyperboloids, paraboloids) 
are common; for instance the mirrors of telescopes generally have rotational symmetry.  
A quadric of revolution is determined by at most six points on it.  
The general equation of the quadratic surfaces is accompanied with certain conditions 
coming from the rotational symmetry, reducing the number of independent constants to six (or less).  
For instance the sphere is determined by four points on it; 
right-circular cone, the right-circular cylinder and the paraboloid of revolution 
are determined by five points each respectively.   
\begin{Theorem}\label{theorem-one}
A quadric is a surface of revolution if and only if it has two equal non-zero 
characteristic roots; {\em i.e.,} the non-zero eigenvalues of the characteristic  
equation~(\ref{quadratic-surfaces-characteristic-equation}) repeat twice.  
\end{Theorem}
This theorem translates into following three possible cases: 

\begin{description}

\item[Case--1:]
None of the mixed-terms constants, $f$, $g$, and $h$ are zero. 

If $fgh \ne 0$, then 
\begin{eqnarray}
\frac{F}{f} = \frac{G}{g} = \frac{H}{h}\,,
\label{case-1}
\end{eqnarray}
where $F$, $G$, and $H$ are the cofactors of $f$, $g$, and $h$ in the 
matrix $e$ ({\em i.e,} $F = gh - af$, $G = hf - bg$, and $H = fg - ch$). 

\item[Case--2:]
If $fgh = 0$, then at least two of the mixed-terms constants, $f$, $g$, and $h$ 
are zero ({\em i.e,} two are necessarily zero and the third one need not be zero).

If $f = 0$, $g = 0$, $h \ne 0$, then 
\begin{eqnarray}
(a - c) (b - c) = h^2\,. 
\label{case-2-a}
\end{eqnarray}

If $g = 0$, $h = 0$, $f \ne 0$, then 
\begin{eqnarray}
(b - a) (c - a) = f^2\,. 
\label{case-2-b}
\end{eqnarray}

If $f = 0$, $h = 0$, $g \ne 0$, then 
\begin{eqnarray}
(a - b) (c - b) = g^2\,. 
\label{case-2-c}
\end{eqnarray}

\item[Case--3:]
When the mixed-terms constants, $f$, $g$, and $h$ are all zero.

If $f = 0$, $g = 0$, and $h = 0$, then 
\begin{eqnarray}
(a - b) (b - c) (c - a) = 0\,, \quad \Rightarrow a = b {\rm ~or~} b = c {\rm ~or~} c = a\,.
\label{case-3}
\end{eqnarray}

\end{description}
Equations~(\ref{case-1})-(\ref{case-3}) summarize the necessary and sufficient conditions 
imposed on the general equation of second degree to represent a surface of 
revolution~\cite{Shanti-Narayan}.  
It is interesting to note that the conditions are expressed in terms of the quadratic 
constants alone and none of the constants of the linear terms ($p$, $q$ and $r$) are used.

\setcounter{section}{0}
 
\section*{}
\addcontentsline{toc}{section}
{Appendix F. \\
Quadricmeter}

\renewcommand{\theequation}{F.{\arabic{equation}}}
\setcounter{equation}{0}
 
\begin{center}
 
{\Large\bf
Appendix F. \\
Quadricmeter \\
(Abstract of the Patent Application)}\footnote{\normalsize
This is the abstract of the patent application as it appeared in: 

\begin{itemize}
\item
Sameen Ahmed Khan, 
{\bf Quadricmeter}, 
{\em Official Journal of the Patent Office}, Issue No. {\bf 43/2008}, Part-I, pp. 25296 (24 October 2008). 
Application No.: {\bf 2126/MUM/2008 A},  
International Classification: {\bf B69G1/36}, 
Controller General of Patents Designs and Trade Marks, Government of India. 
\url{http://ipindia.nic.in/ipr/patent/journal_archieve/journal_2008/patent_journal_2008.htm},  
\url{http://ipindia.nic.in/ipr/patent/journal_archieve/journal_2008/pat_arch_102008/official_journal_24102008_part_i.pdf}, 
\url{http://www.ipindia.nic.in/}, 
({\em patent in process}, \url{http://SameenAhmedKhan.webs.com/quadricmeter.html}). 
\end{itemize}
See the references~\cite{quadricmeter, generalized-spherometer} for details and updates. 
}

\end{center}

\begin{abstract}
{\bf Quadricmeter} is the instrument devised to identify (distinguish) and measure the various parameters 
(axis, foci, latera recta, directrix, etc.,) completely characterizing the important class of surfaces 
known as the quadratic surfaces.  Quadratic surfaces (also known as quadrics) include a wide range of 
commonly encountered surfaces including, cone, cylinder, ellipsoid, elliptic cone, elliptic cylinder, 
elliptic hyperboloid, elliptic paraboloid, hyperbolic cylinder, hyperbolic paraboloid, paraboloid, 
sphere, and spheroid.  Quadricmeter is a generalized form of the conventional spherometer and the 
lesser known cylindrometer (also known as the {\em Cylindro-Spherometer} and {\em Sphero-Cylindrometer}).  
With a conventional spherometer it was possible only to measure the radii of spherical surfaces.  
Cylindrometer can measure the radii of curvature of a cylindrical surface in addition to the spherical 
surface.  In both the spherometer and the cylindrometer one assumes the surface to be either spherical 
or cylindrical respectively.  In the case of the quadricmeter, there are no such assumptions. 
\end{abstract}

\setcounter{section}{0}
 
\section*{}
\addcontentsline{toc}{section}
{Appendix G. \\
Optical Techniques}

\renewcommand{\theequation}{G.{\arabic{equation}}}
\setcounter{equation}{0}
 
\begin{center}
 
{\Large\bf
Appendix G. \\
Optical Techniques
} \\

\end{center}

Apart from the contact techniques, it is also possible to use the optical techniques.   
The major source of inaccuracy in using a spherometer (and its variants describes in this article) 
is in determining the exact point of contact between the probe and the surface being tested.  
Sharper legs of the spherometer can damage the surface and this can be circumvented to a good extent 
by the use of the ball-spherometer (and the ball-cylindrometer suggested by the author in this note).  
This necessitates the creation of non-contact methods. 

Optical techniques based on the interference patterns provide one such means of 
non- contact procedure to study the surfaces~\cite{Cylindrometer-In-House}. 
In some spherometers the point of contact is determined by observing the Newton's ring
interference pattern formed between the test surface and an optical surface mounted on
the end of the probe.  As the probe is brought up to the surface, the ring pattern expands,
but when the point of contact is reached, no further motion occurs.  
Interferometry is a precise and non-destructive tool and determines other optical parameters 
(such as thickness, refractive index) in addition to the radius of curvature.  
Interferometer are also used for measuring surface flatness of plano optical elements such as 
mirrors, prisms and windows.  

A Fizeau interferometer can also be used to measure radius of curvature of a surface.  A
surface having a long radius of curvature can be compared interferometrically with a flat
surface to yield Newton's rings.  
The Fizeau interferometer is the most commonly used interferometer for testing optical 
components and systems used aboard space borne or space-related instrumentation.  

Non-contacting and whole-field method of measuring curvatures using a moving laser source 
and an image-shearing camera have been also developed.  
There are also procedures for the surface reconstruction based on the 
transmitted wavefront.  The reconstructed surface is a complete surface 
including its radius of curvature~\cite{Selberg, Abdelsalama}.

\setcounter{section}{0}
 
\section*{}
\addcontentsline{toc}{section}
{Appendix H. \\
Radius of Curvature}

\renewcommand{\theequation}{H.{\arabic{equation}}}
\setcounter{equation}{0}
 
\begin{center}
 
{\Large\bf
Appendix H. \\
Radius of Curvature
} \\

\end{center}
 
Curvature is the amount by which a geometric object deviates from being flat, 
or straight in the case of a line~\cite{Weisstein-Radius, Gray-Radius}.  
The curvature $\kappa$ of a circle is defined as the reciprocal of its radius $R$.  
Smaller circles bend more sharply, and hence have higher curvature.  
The circles and spheres are unique as they have a constant curvature. 
The curvature of a smooth curve is defined as the curvature of its {\em osculating circle} 
at each point.
Let $y = f(x)$, then the radius of curvature is given by 
\begin{eqnarray}
R = \frac{\left[1 + \left( \frac{d y}{d x} \right)^2 \right]^{3/2}}
{\left| \frac{d^2 y }{d x^2} \right|}\,.  
\label{radius-of-curvature}
\end{eqnarray}
For example, in an ellipse with major axis $2a$ and minor axis $2b$, the vertices 
on the major axis have the smallest radius of curvature of any points $R = {b^2}/a$ 
and the vertices on the minor axis have the largest radius of curvature of any points 
$R = {a^2}/b$.  

If the curve is given parametrically by the functions $x (t)$ and $y (t)$, then the 
radius of curvature is
\begin{eqnarray}
R = \frac{\left[\dot{x}^2 + \dot{y}^2 \right]^{3/2}}
{\left| \dot{x} \ddot{y} - \dot{y} \ddot{x} \right|}\,,  
\label{radius-of-curvature-parametric}
\end{eqnarray}
where $\dot{x} = {d x}/{d t}$, $\ddot{x} = {d^2 x}/{d t^2}$, $\dot{y} = {d y}/{d t}$ 
and $\ddot{y} = {d^2 x}/{d t^2}$.

\end{document}